\documentclass[twocolumn,aps,superscriptaddress]{revtex4}

\usepackage{amssymb}
\usepackage{amsmath}
\usepackage{graphicx}
\usepackage[normalem]{ulem}
\usepackage[dvips]{color}

\begin{document}

\title{Charge asymmetry dependence of the elliptic flow splitting in relativistic heavy-ion collisions}
\author{Zhang-Zhu Han}
\affiliation{Shanghai Institute of Applied Physics, Chinese Academy
of Sciences, Shanghai 201800, China}
\affiliation{University of Chinese Academy of Sciences, Beijing 100049, China}
\author{Jun Xu\footnote{corresponding author: xujun@sinap.ac.cn}}
\affiliation{Shanghai Advanced Research Institute, Chinese Academy of Sciences,
Shanghai 201210, China}
\affiliation{Shanghai Institute of Applied Physics, Chinese Academy
of Sciences, Shanghai 201800, China}

\date{\today}

\begin{abstract}
The elliptic flow splitting $\Delta v_2$ between $\bar{u}$ and
$u$ quarks as well as between $\pi^-$ and $\pi^+$ in midcentral Au+Au collisions at $\sqrt{s_{NN}}=200$ GeV has been studied, based on
the framework of an extended multiphase transport model with the partonic evolution
described by the chiral kinetic equations of motion. Within the available statistics, the slope of $\Delta v_2$  between $\bar{u}$ and
$u$ quarks with respect to the electric charge asymmetry $A_{ch}$ from the linear fit is found to be negative, due to the correlation between
the velocity and the coordinate in the initial parton phase-space distribution. Simulations with the magnetic field in QGP overestimate the splitting of the spin polarization between $\Lambda$ and $\bar{\Lambda}$ observed experimentally, with the latter more consistent with results under the magnetic field in vacuum. Considering the uncertainties from the magnetic field, the quark-antiquark vector interaction, and the hadronization, as well as the hadronic evolution, our study shows that the experimentally observed positive slope of $\Delta v_2$ with respect to $A_{ch}$
is not likely due to the chiral magnetic wave.
\end{abstract}

\maketitle

\section{INTRODUCTION}
\label{introductionhan}

Understanding the properties of the quark-gluon plasma (QGP) is one of the main goals of relativistic heavy-ion experiments. Due to the high
energy density reached in relativistic heavy-ion collisions, the chiral symmetry of quarks is restored, and they can be considered approximately
as massless particles with chiralities. With the magnetic field mainly generated by spectator protons in noncentral relativistic heavy-ion
collisions, various interesting phenomena as a result of the chiral dynamics of these partons can be observed. For example, a net axial charge
current along the direction of the magnetic field can be produced from a net electric charge density, while a net electric charge current along
the direction of the magnetic field can be produced from a net axial charge density. The former is called the chiral separation
effect~\cite{Son04,Metli05,Son09} and the latter is called the chiral magnetic effect~\cite{Khar08,Fuku082,Khar10}. Their interplay leads
to the chiral magnetic wave (CMW)~\cite{Khar11}, a wave mode with the axial and the electric charges oscillating in space and time.

It was proposed in Ref.~\cite{Burn11} that the CMW can lead to the electric quadrupole moment of the QGP formed in the transverse plane in noncentral relativistical heavy-ion collisions. The latter may result in the elliptic flow splitting between particles of opposite electric charges. As the effect of the CMW depends on the magnitude of the electric charge chemical potential, the elliptic flow splitting is expected to be larger with a larger charge asymmetry. The positive slope of the elliptic flow difference between $\pi^-$ and $\pi^+$ with respect to the electric charge asymmetry has been experimentally observed~\cite{STAR15}. In order to study the chiral dynamics and understand the experimental results, various hydrodynamic models have been developed by incorporating the chiral anomalies~\cite{Hydro1,Son09,Liao}. In addition, the chiral kinetic equations of motion describing the anomalous transport of massless fermions have been derived from different approaches~\cite{Step12,Son13,Chen13,Man14,Zhou18}, making it possible to study the CMW and the resulting elliptic flow splitting of charged particles from transport simulations in relativistic heavy-ion collisions. From an initially thermalized parton distribution and a parameterized decaying magnetic field, it was found that the slope of the elliptic flow splitting with respect to the electric charge asymmetry is negative with the dynamics described by the full chiral kinetic equations of motion~\cite{Sun16}, while this is not much affected by the vorticity of the system~\cite{Sun171}.

How large the CMW effect is depends on the strength of the magnetic field, with the latter leading to the splitting of the spin polarization between particles and their antiparticles. It was found that the spin polarizations of $\Lambda$ and $\bar{\Lambda}$ are small at $\sqrt{s_{NN}}=200$
GeV~\cite{STAR07,STAR18}, while they become larger at lower collision energies~\cite{Lisa16} as a result of the stronger
stopping and the higher angular velocity. The flow vorticity and the resulting polarization were predicted in Refs.~\cite{Cse13,Bec13,Bec15,Xie16}, and the later studies~\cite{Ari16,Pang16,Kar17,LiHui17,Sun17,Li17,Baz18} showed quantitatively that the vorticity field leads to the same $\Lambda$ and $\bar{\Lambda}$ spin polarization. The splitting of their spin polarizations can be induced by the magnetic field~\cite{Mul18} and the vector strong interaction~\cite{Han18,Cse19}. At $\sqrt{s_{NN}}=200$ GeV, we found that the splitting of the spin polarization between $\Lambda$ and $\bar{\Lambda}$ is mostly determined by the magnetic field. It will be useful to consider together the electric charge asymmetry dependence of the elliptic flow splitting and the splitting of the spin polarization between $\Lambda$ and $\bar{\Lambda}$, since they are potentially induced by the magnetic field but are independent of the flow vorticity.

In the present study, we do a state-of-the-art transport simulation in order to understand the elliptic flow splitting of charged particles
induced by the CMW in noncentral Au+Au collisions at $\sqrt{s_{NN}}=200$ GeV based on the framework of an extended multiphase transport (AMPT)
model. Typically, the initial parton distribution is taken from that of the AMPT model, and the space-time evolution of the magnetic field generated by
spectator protons is calculated by considering the QGP response. The effects of the quark-antiquark vector interaction and the hadronization as
well as the hadronic evolution on the elliptic flow splitting are also investigated. The electric charge asymmetry dependence in the partonic system and the hadronic system are both analyzed in a way similar to the experimental analysis. With all these effects incorporated properly in the transport simulation, our conclusion turns out to be that the positive slope of the elliptic flow difference between $\pi^-$ and $\pi^+$ with respect to the electric charge asymmetry observed experimentally is not likely due to the CMW.

The rest part of the paper is organized as follows. In Sec.~\ref{theoryhan}, we give a brief review of the extended AMPT model, the chiral kinetic equations of motion, and the calculation of the magnetic field. Section~\ref{resultshan} gives the detailed results of the elliptic flow splitting between negative and positive charged particles with respect to the electric charge asymmetry, as well as the spin polarizations of different quark species. A summary is given in Sec.~\ref{summaryhan}.

\section{Theoretical framework}
\label{theoryhan}

\subsection{The extended AMPT model}

The present study is based on an extended AMPT model by mainly incorporating the chiral dynamics in the partonic phase~\cite{Han18}, as an extension
of the string melting version of the original AMPT model~\cite{Lin05}. The momenta of initial partons are from melting hadrons produced by the
heavy ion jet interaction generator (HIJING) model~\cite{Wang91}. Their coordinates in the transverse plane $(x, y)$ are set to be the same as
those of the colliding nucleons that produce their parent hadrons, while the longitudinal coordinates are almost 0 due to the strong Lorentz
contraction at $\sqrt{s_{NN}}=200$ GeV. Each parton is given a formation time
related to the energy and the transverse mass of its parent hadron~\cite{Lin05}. The dynamics of these partons is described by the chiral kinetic
equations of motion under the magnetic field mainly generated by spectator protons to be detailed in the next subsections. The partonic dynamics
also includes two-body scatterings with an isotropic cross section of 1.5 mb~\cite{Xu11}, and it ends at around $t \sim 3.7$ fm/c when the
partonic scatterings are almost finished. Afterwards, a spatial coalescence scenario is used to model the hadronization process, which allows a
pair of nearest quark and antiquark to form a meson and three nearest quarks (antiquarks) to form a baryon (antibaryon), with the mass and
species of the hadron determined by the invariant mass and the flavors of the constituent partons. The hadronic evolution is described by a
relativistic transport (ART) model~\cite{Li95} that contains various elastic, inelastic, and decay channels. The electric charge is conserved in inelastic and decay channels in the present version of AMPT/ART~\cite{Xu16}.

\subsection{The chiral kinetic equations of motion}

For the partonic dynamics, we start from the Lagrangian of the 3-flavor Nambu-Jona-Lasinio (NJL) model under an external electromagnetic field
\begin{eqnarray}\label{originalLag}
\mathcal{L}&=&\bar{\psi}\gamma_\mu(i\partial^\mu - QA_{ext}^{\mu})\psi - \bar{\psi} M \psi \nonumber \\
&+&\frac{G_{S}}{2}\sum\limits_{\alpha=0}^8 [(\bar{\psi}\lambda^{\alpha}\psi)^{2}+(\bar{\psi}i\gamma_{5}\lambda^{\alpha}\psi)^{2}] \nonumber \\
&-&\frac{G_{V}}{2}\sum\limits_{\alpha=0}^8 [(\bar{\psi}\gamma_{\mu}\lambda^{\alpha}\psi)^{2}
+(\bar{\psi}\gamma_{\mu}\gamma_{5}\lambda^{\alpha}\psi)^{2}] \nonumber \\
&-&K\{{\rm det}_f[\bar{\psi}(1+\gamma_{5})\psi]+{\rm det}_f[\bar{\psi}(1-\gamma_{5})\psi]\},
\end{eqnarray}
where $\psi=(\psi_{u},\psi_{d},\psi_{s})^{T}$ is respectively the quark field for $u$, $d$, and $s$
quarks, $Q={\rm diag}(q_{u}e,q_{d}e,q_{s}e)$ represents respectively their electric charges, $A_{ext}^{\mu}=(\varphi,\vec{A}_{m})$ is the
external electromagnetic potential, $M={\rm diag}(m_{u},m_{d},m_{s})$ is the current quark mass matrix,
$\lambda^{\alpha}$ are the Gell-Mann matrices in SU(3) flavor space
with $\lambda^{0}=\sqrt{{2}/{3}}I$, and $G_{S}$ and $G_{V}$ are,
respectively, the scalar and the vector coupling constant. The value of the vector coupling constant $G_{V}$ affects the critical point of the
chiral phase transition in the phase diagram~\cite{As89,Fuku081,Car10,Bra13}. The value of $G_V$ is chosen to be 0 or $1.1G_S$~\cite{Lutz92} in
the present study, in order to account for its large uncertainty. The $K$ term, with ${\rm det}_{f}$ denoting the determinant in the flavor space,
is the Kobayashi-Maskawa-'t Hooft interaction~\cite{Hooft78} which breaks the axial $U_{A}(1)$ symmetry.

This model describes reasonably well the chiral symmetry restoration of quarks at high temperatures at the early stage of the partonic dynamics,
when quarks with very small dynamical masses can be approximately considered as massless particles. This approximation becomes worse if partons
are about to freeze out, when the strength of the magnetic field is expected to be rather weak. Taking the mean-field approximation and in the
limit of massless partons, the Lagrangian can be approximately written as
\begin{equation}\label{simpleLag}
\mathcal{L}_{MF}\approx\bar{\psi}\gamma_{\mu}(i\partial^{\mu}-QA_{ext}^{\mu}-g_{V}\langle\bar{\psi}\gamma^{\mu}\psi\rangle)\psi,
\end{equation}
with $g_V=\frac{2}{3}G_{V}$ and the corresponding term representing the
flavor-singlet quark-antiquark vector interaction from the mean-field approximation~\cite{As89,Hat94}. The quark-antiquark vector density in
Eq.~(\ref{simpleLag}) can be expressed as
\begin{equation} \label{vectordensity}
\langle\bar{\psi}\gamma^{\mu}\psi\rangle=2N_{c}\sum\limits_{i=u,d,s} \int\frac{d^{3}k}{(2\pi)^{3}E_{i}}k^{\mu}(f_{i}-\bar{f}_{i}),
\end{equation}
where $N_{c}=3$ is the color degeneracy, $E_{i}=k$ is the energy for massless quarks (antiquarks), and
$f_{i}$ and $\bar{f}_{i}$ are respectively the phase-space distribution functions of quarks and
antiquarks of flavor $i$, which are calculated from the test-particle method~\cite{Won82,Ber88} by averaging over
parallel events in the transport simulation. As in the original NJL model, the above momentum
integration is cut off at $750$ MeV/c~\cite{Lutz92,Bra13}.

The Euler-Lagrange equation for quark flavor
$i$ can be obtained from the Lagrangian after the mean-field approximation [Eq.~(\ref{simpleLag})] as
\begin{equation}\label{euler}
[\gamma^{\mu}(i\partial_{\mu}-A_{\mu})]\psi_{i}=0.
\end{equation}
In the above, $A_{\mu}=(A_{0},-\vec{A})$ contains the time and space
components of the vector potential. We neglect the time component $A_0$ in the following, since the purpose of the present study is to
investigate the effect of the CMW induced by the space component of the vector potential $\vec{A}$, which is expressed as
\begin{eqnarray}
\vec{A}&=&b_{i}g_{V}\vec{\rho}+q_{i}e\vec{A}_{m}, \label{totala}
\end{eqnarray}
with $b_{i}=1$ for quarks and $-1$ for antiquarks being the baryon charge number, and $q_{i}$ being
the electric charge number of the quark species $i$. $\vec{\rho}\equiv\langle\bar{\psi}\vec{\gamma}\psi\rangle$ is the space component of the
quark-antiquark vector density, and $\vec{A}_{m}$ is the vector potential of the magnetic field generated by spectator protons to be detailed in
the next subsection.

After decoupling the $4\times4$ Euler-Lagrange equation [Eq.~(\ref{euler})] into the $2\times2$ Schr\"odinger equation, the
single-particle Hamiltonian can be obtained as
\begin{equation}\label{hamiltonian}
H=c\vec{\sigma}\cdot\vec{k},
\end{equation}
where $\vec{k}=\vec{p}-\vec{A}$ is the real momentum and $\vec{p}$ is
the canonical momentum of the particle. $c=\pm 1$ is the helicity of the particle, and
$\vec{\sigma}$ are the Pauli matrics. In the semiclassical picture, the evolution of the system from the above Hamiltonian can be described as
\begin{eqnarray}
\dot{\vec{r}}&=&c\vec{\sigma}, \label{rdot} \\
\dot{\vec{k}}&=&c\vec{\sigma}\times\vec{B}, \label{kdot} \\
\dot{\vec{\sigma}}&=&2c\vec{k}\times\vec{\sigma},    \label{sigmadot}
\end{eqnarray}
where $\vec{B}=\nabla\times\vec{A}$
is the total magnetic field, including the
contributions from the real magnetic field $\vec{A}_{m}$ generated by spectator
protons and the effective magnetic field originated from the net quark flux $\vec{\rho}$, with the former acting on the electric charge and the
latter acting on the baryon charge of the particle according to Eq.~(\ref{totala}). Using the adiabatic approximation $\vec{\sigma}\approx
c\hat{k}-\frac{\hbar}{2k}\hat{k}\times\dot{\hat{k}}$
that satisfies $\hat{k}\cdot \dot{\vec{r}}\approx 1+O(\hbar^{2})$, the chiral kinetic
equations of motion under an external magnetic field can be obtained as~\cite{Step12,Son13,Chen13,Man14,Zhou18}
\begin{eqnarray}
\sqrt{G}\dot{\vec{r}}&=&\hat{k}+\frac{c\hbar}{2k^{2}}\vec{B}, \label{rdot1} \\
\sqrt{G}\dot{\vec{k}}&=&\hat{k}\times\vec{B},\label{kdot1}
\end{eqnarray}
with $\sqrt{G}=1+c\hbar\vec{B}\cdot\vec{k}/(2k^{3})$. In the present study, the contribution of the electric field in the above chiral kinetic equations of motion is neglected. Starting from the
initial phase-space distribution from the AMPT/HIJING model
and with equal numbers of partons having positive ($c=1$) and negative ($c=-1$) helicity,
the partonic phase evolves according to the above chiral kinetic equations of motion.
The $\sqrt{G}$ factor in Eq.~(\ref{kdot1}) leads
to the modification of the phase-space volume~\cite{Xiao05}, so the statistical value of
any observable $X$ is calculated according to
$\langle X \rangle=\sum_{n}\sqrt{G_{n}}X_{n}/\sum_{n}\sqrt{G_{n}}$ by
taking $\sqrt{G_n}$ of the $n$th particle as a weight factor.

\subsection{The magnetic field in vacuum and in QGP}

The vector potential of the real magnetic field $\vec{A}_{m}$ in vacuum can be expressed as
\begin{eqnarray}
\vec{A}_{m}(t,\vec{r})&=&\frac{e}{4\pi}\sum\limits_{n} Z_{n}\frac{\vec{v}_{n}}{R_{n}-\vec{v}_{n}\cdot\vec{R}_{n}}, \label{mpinvacuum}
\end{eqnarray}
where $Z_{n}$ is the charge number of the $n$th spectator nucleon,
$\vec{v}_{n}$ is the velocity of the $n$th spectator nucleon at the retarded
time $t'_{n}=t-|\vec{r}-\vec{r}_{n}|$ when the spectator nucleon emits
radiation, and $\vec{R}_{n}=\vec{r}-\vec{r}_{n}$ is the relative
position of the field point $\vec{r}$ with respect to the spectator nucleon
position $\vec{r}_{n}$. Considering the finite electrical conductivity of the QGP,
the vector potential of the magnetic field induced by a point particle with
charge $e$ moving in the $+z$ direction at the velocity $v$ along the trajectory
$z=vt+z_{0}$ is expressed as~\cite{Tuchin16}
\begin{eqnarray}\label{mpinplasma}
\vec{A}_{m}^{e}&=&\frac{\hat{z}e}{4\sigma_{con}[(z-z_{0})/v]}\times\frac{\exp\left\{\frac{-b^{2}}{4\{\lambda(t)-\lambda[(z-z_{0})/v]\}}\right\}}{4\{\lambda(t)-\lambda[(z-z_{0})/v]\}}
\nonumber  \\
&\times&\theta[vt-(z-z_{0})]\theta[(z-z_{0})-vt_{0}] \nonumber \\
&+&\frac{\hat{z}ev\gamma}{4\pi}\int_{0}^{+\infty}dk_{\perp}J_{0}(k_{\perp}b) \nonumber\\
&\times&\exp[-k_{\perp}^{2}\lambda(t) -k_{\perp}\gamma|(z-z_{0})-vt_{0}|],
\end{eqnarray}
where $t_{0}$ is the time when the QGP emerges, $\sigma_{con}(t)$ is the electrical conductivity of
the QGP with $\lambda(t)=\int_{t_{0}}^{t}dt^{'}/\sigma_{con}(t^{'})$ related to its time evolution,
$\gamma=1/\sqrt{1-v^{2}}$ is the Lorentz factor and $b$ is the distance between the field point and the
point particle with charge $e$ perpendicular to the $z$ direction, $J_{0}$ is the zeroth-order Bessel
function of the first kind, and $\theta$ is the Heaviside step function. As mentioned in Ref.~\cite{Tuchin16}, Eq.~(\ref{mpinplasma}) is valid only for a spatially uniform electrical conductivity. Equation~(\ref{mpinvacuum}) is used to
calculate $\vec{A}_{m}$ in vacuum before $t_{0}$,
and Eq.~(\ref{mpinplasma}) is used to calculated $\vec{A}_{m}$ from the summation of
$\vec{A}_{m}^{e}$ after $t_{0}$ when the QGP is produced. Since partons are continuously produced and
$\sigma_{con}$ increases gradually from 0 to finite, $t_{0}$ should in principle be set as early as
possible. In the present study we choose $t_{0}\sim 0.09$ fm/c, before which there are too few partons
leading to large fluctuations.

\section{Results and discussions}
\label{resultshan}

Based on the framework of the extended AMPT model, we simulate the dynamics according to the chiral kinetic equations of motion under the
magnetic field. Using the same chiral kinetic equations of motion and under a constant external magnetic field, the CMW is simulated in
Ref.~\cite{Zhou18} and the electric quadrupole moment can be observed in a thermalized box system with the periodic boundary condition. Starting
from a thermalized almond-shaped QGP medium and under a parameterized decaying magnetic field, the electric quadrupole moment generated at the
freeze-out stage of the partonic phase by the CMW can also be observed, as shown in
Ref.~\cite{Sun16}. In the present study, the simulation is done from a more realistic initial parton distribution and under a more realistic
space-time evolution of the magnetic field. We will see how the quark-antiquark vector interaction, the hadronization, and the hadronic evolution
affect the elliptic flow splitting between opposite charged particles with respect to the electric charge asymmetry of the system. In addition, the splitting of the spin polarizations between quarks of opposite charges serving as a probe of the space-time evolution of the magnetic field is also studied.

\subsection{Space-time evolution of the magnetic field}

\begin{figure}[ht]
	\includegraphics[scale=0.3]{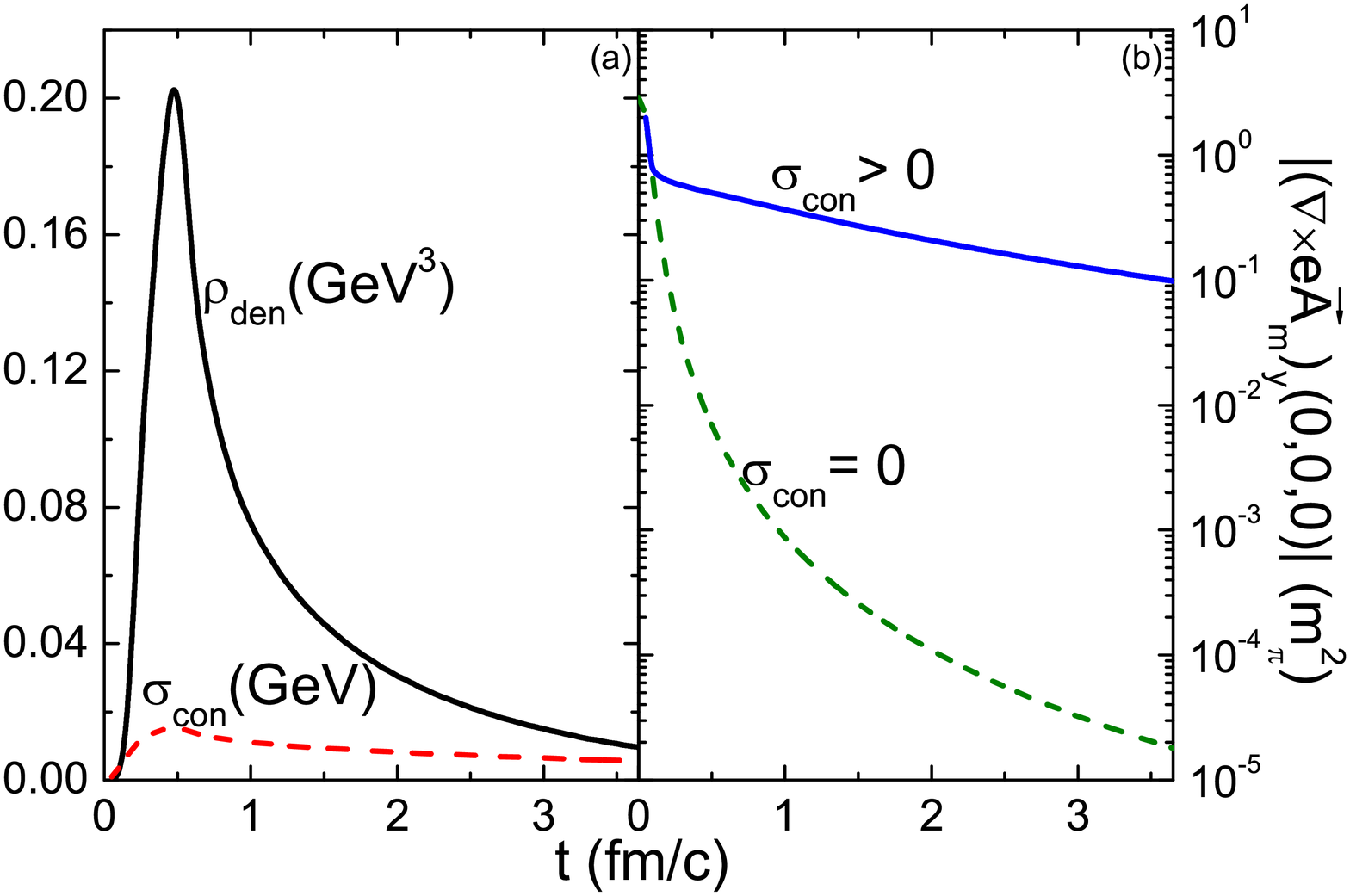}
	\caption{(Color online) Left: Time evolution of the total light parton number density in central cells and the approximated electrical
conductivity of the QGP; Right: Time evolution of the $y$ component of the real magnetic field in vacuum ($\sigma_{con}=0$) and in QGP
($\sigma_{con}>0$) at the center of the collision system. Results are from simulations for midcentral Au+Au collisions at $\sqrt{s_{NN}}=200$
GeV.} \label{Fig1}
\end{figure}

We first display how the magnetic field evolves with time in the present study. Figure~\ref{Fig1} (a) displays the time evolution of the total light
parton number density $\rho_{den}$ in central cells and the approximated electrical conductivity $\sigma_{con}$ of the quark matter formed in
midcentral ($20-50\%$) Au+Au collisions with an average impact parameter $8.87$ fm at
$\sqrt{s_{NN}}=200$ GeV from transport simulations. The calculation of the electrical conductivity of the QGP is quite model-dependent~\cite{Arn00,Arn03,Gup04,Aar07,Bui10,Bur12,Bra13a,Ama13,Cas13,Ste14,Sah18}, and here we take the temperature dependence of the electrical
conductivity from the lattice QCD result as~\cite{Ding11}
\begin{eqnarray}
\sigma_{con}=0.0058\frac{T}{T_{c}}({\rm GeV}), \label{conductivity}
\end{eqnarray}
where $T$ is the temperature of the QGP and $T_{c}\approx0.165$ GeV is the critical temperature. Here we assume the extreme case that $\sigma_{con}$ is uniform in QGP as in Ref.~\cite{McL14}, while we take into account the time evolution of $\sigma_{con}$ as a result of the gradual formation of QGP as well as the expansion of the system, going beyond the approximations made in Ref.~\cite{McL14}. The time evolution of the temperature $T$ in Eq.~(\ref{conductivity}) is taken from the central cells of the QGP£¬ which is a thermalized system consisting massless particles with their momenta
following approximately the Boltzmann distribution. The temperature $T$ is extracted from the light quark density $\rho_{den}$ there through the
relation $T\approx(\pi^{2}/24)^{1/3}\rho_{den}^{1/3}$. Figure~\ref{Fig1} (b) displays the time evolution of the $y$ component of the real
magnetic field in vacuum ($\sigma_{con}=0$) and in QGP ($\sigma_{con}>0$) at the center of the same collision system. It is seen that the magnetic field drops by about a factor of $5\sim10$ in the first $0.15\sim0.20$ fm/c, which is a characteristic time scale of the nucleus thickness in the beam direction. This feature is discussed in Ref.~\cite{McL14} and also observed in Ref.~\cite{Vor11}. In the later stage, the real magnetic field in vacuum decreases dramatically with time, while the lifetime of the real magnetic field in QGP lasts much longer.

\begin{figure}[ht]
	\includegraphics[scale=0.3]{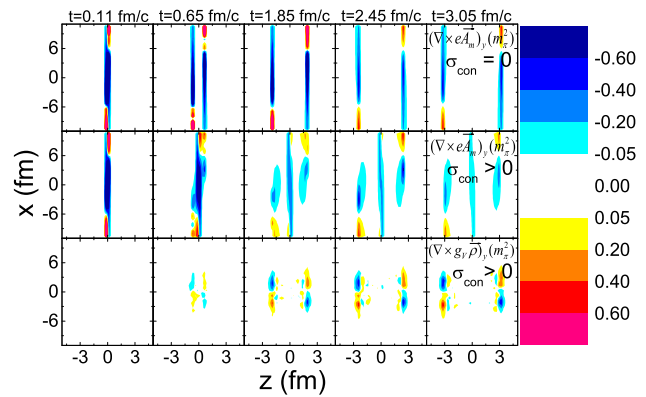}
	\caption{(Color online) Contours of the $y$ component of the real magnetic field $(\nabla\times
		e\vec{A}_{m})_{y}$ in vacuum (first row), the real magnetic field $(\nabla\times
		e\vec{A}_{m})_{y}$ in QGP (second row), and the effective
		magnetic field $(\nabla\times g_{V}\vec{\rho})_{y}$ (third row) at different times
		in the reaction plane of midcentral Au+Au collisions at $\sqrt{s_{NN}}=200$ GeV.} \label{Fig2}
\end{figure}

Figure~\ref{Fig2} displays the spatial distributions of various fields at different times in the reaction
plane (x-o-z plane) in midcentral Au+Au collisions at $\sqrt{s_{NN}}=200$ GeV. As shown in the first
row, the real magnetic field in vacuum ($\sigma_{con}=0$) in the central region of the
collision system, which is in the $-y$ direction, decreases dramatically with time, so it is expected that its effect on the dynamics of midrapidity particles decays rather rapidly, although the areas where it is strong move with the spectators towards $\pm z$ directions. The spatial distribution of the real magnetic field in the central region in the x-o-z plane is similar to that shown by Fig.~3 of Ref.~\cite{Vor11}. The real magnetic field in QGP ($\sigma_{con}>0$) becomes more diffusive especially in the central region, while the space-time distribution looks similar compared to that in vacuum, as shown in the second row. It is seen that the magnetic field can be in the $+y$ direction only in the regions on the left-hand side of the spectators. Although the net quark density is low at $\sqrt{s_{NN}}=200$ GeV, the net quark flux is not negligible. Compared with the real magnetic field, the effective magnetic field $(\nabla\times g_{V}\vec{\rho})_{y}$ with $G_V=1.1G_S$ from the curl of the net quark flux shows a
completely different space-time distribution, which is positive (negative) at $x\cdot z>0$ ($x\cdot
z<0$), as shown in the third row of Fig.~\ref{Fig2}.

\subsection{Event distribution of the electric charge asymmetry}

\begin{figure}[ht]
	\includegraphics[scale=0.3]{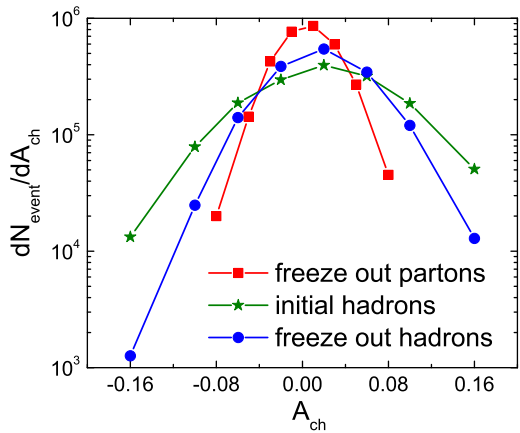}
	\caption{(Color online) Event distributions of the electric charge asymmetry $A_{ch}$
		for freeze-out partons, initial hadrons right after hadronization, and freeze-out hadrons after hadronic evolution in midcentral Au+Au
collisions at $\sqrt{s_{NN}}=200$ GeV.} \label{Fig3}
\end{figure}

Before studying the elliptic flow splitting, we show in Fig.~\ref{Fig3} the event distributions of the electric charge asymmetry at different stages of
midcentral Au+Au collisions at $\sqrt{s_{NN}}=200$ GeV. The electric charge asymmetry is defined as $A_{ch}=\sum_n q_n/\sum_n |q_n|$, where $q_n$ is the electric charge number of the $n$th particle at midrapidities. This definition is consistent with that in the experimental analysis~\cite{STAR15} for freeze-out hadrons. For freeze-out partons or initial hadrons with the charge number other than $\pm 1$, this definition is related to the charge chemical potential for midrapidity particles, which is one of the key source of the CMW. It is seen that the average value of $A_{ch}$ is consistent with 0 or a very small positive value at all stages in midcentral Au+Au collisions at $\sqrt{s_{NN}}=200$ GeV, and it is generally
asymmetrically distributed at the two sides of the average value. Similar to the experimental analysis~\cite{STAR15}, we will study the slope of
the elliptic flow splitting between negative and positive charged particles with respect to $A_{ch}$ around its average value. It is also
interesting to see that the freeze-out partons have a narrow $A_{ch}$ event distribution due to the large multiplicity, while the distribution
becomes wider for initial hadrons right after the hadronization as a result of the reduction of the particle multiplicity. Due to the increasing
number of hadrons during the hadronic evolution, the $A_{ch}$ event distribution for freeze-out hadrons becomes narrow again.

\subsection{Charge asymmetry dependence of the elliptic flow splitting in the partonic phase}

\begin{figure}[ht]
	\includegraphics[scale=0.265]{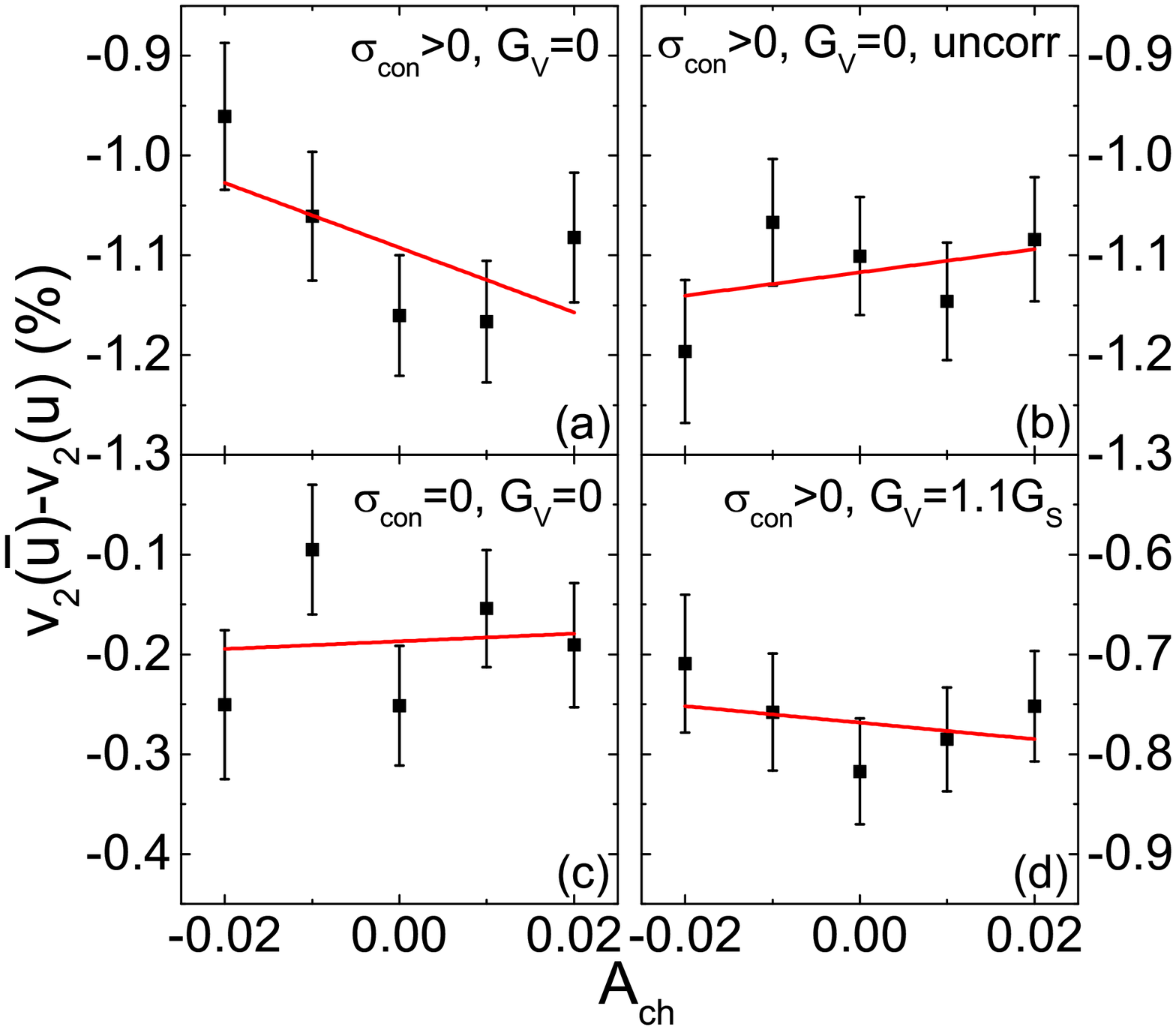}
	\includegraphics[scale=0.25]{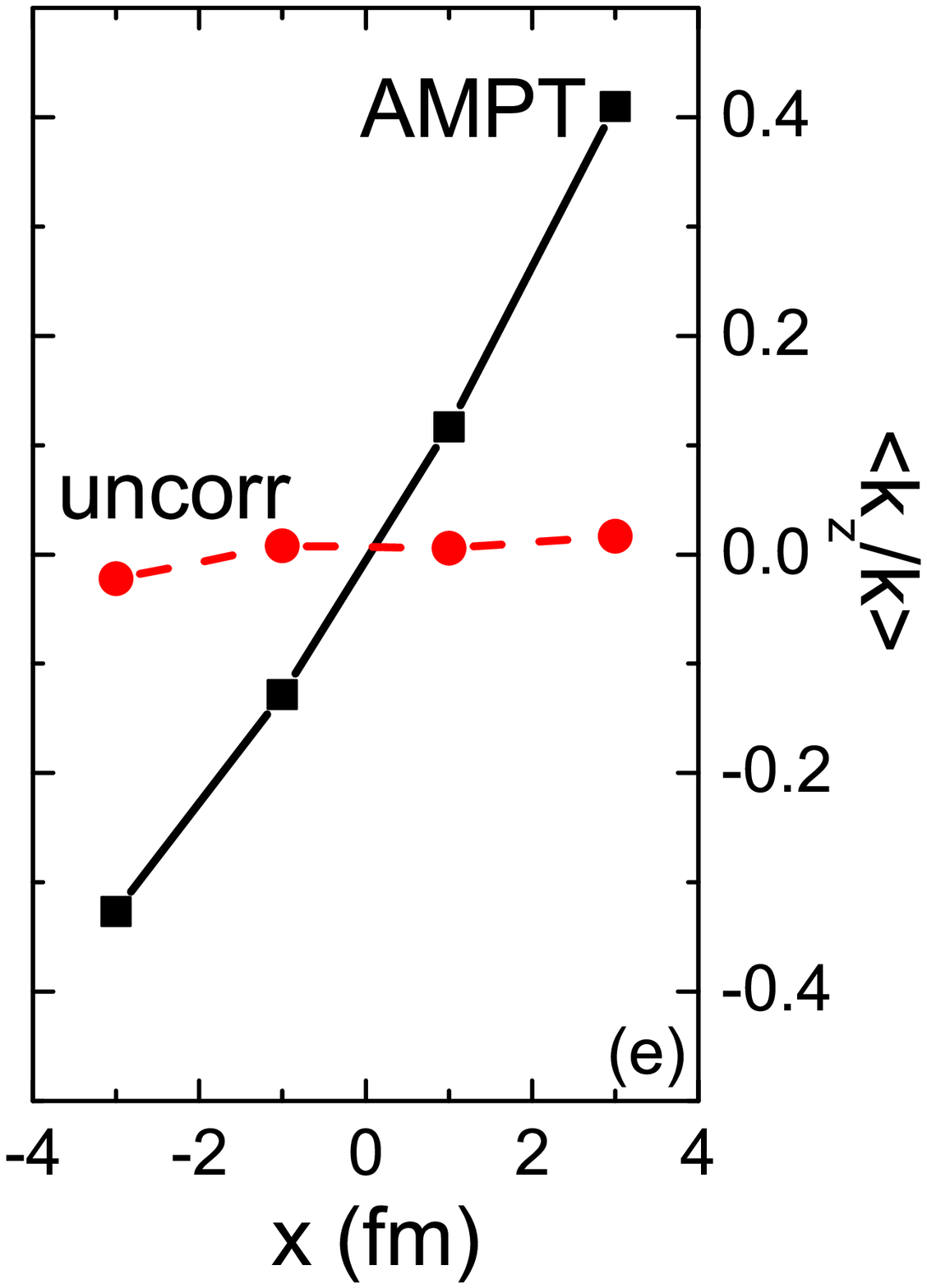}
	\caption{(Color online) Left: Elliptic flow difference between freeze-out $\bar{u}$ and $u$ quarks
    as a function of the electric charge asymmetry $A_{ch}$ in midcentral Au+Au collisions at
	$\sqrt{s_{NN}}=200$ GeV for different scenarios.
    Right: Correlated relation between the $z$-component velocity $\langle k_{z}/k\rangle$ and the coordinate $x$ for initial quarks from AMPT
    and the uncorrelated one for testing purposes.} \label{Fig4}
\end{figure}

With different scenarios of the partonic evolution, we compare the elliptic flow ($v_2$) difference between freeze-out $\bar{u}$ and $u$ quarks
as a function of the electric charge asymmetry $A_{ch}$ of freeze-out partons in the left panel of Fig.~\ref{Fig4}, and partons in the
transverse momentum range $0.15<p_{T}<0.5$ GeV/c at midrapidities are counted for analysis. The
results are based on about 220,000 events for each scenario. The overall larger $\bar{u}$ elliptic
flow than $u$ quarks is observed as a result of neglecting the time component of the vector potential, while in the present study we focus on the
slope of the $v_2$ difference with respect to $A_{ch}$. Considering the magnetic field in QGP but without the quark-antiquark vector interaction
($\sigma_{con}>0$ and $G_{V}=0$), a negative slope of $v_2(\bar{u})-v_2(u)$ with respect to $A_{ch}$ is observed, consistent with that observed
in Ref.~\cite{Sun16} but opposite to the corresponding slope observed experimentally~\cite{STAR15}. This is due to the correlation between the
$z$-component velocity and the $x$ coordinate in the initial parton distribution from the AMPT/HIJING model, which is consistent with the initial
vorticity of the system perpendicular to the reaction plane. With such a correlation as shown in the right panel of Fig.~\ref{Fig4}, the Lorentz force
in Eq.~(\ref{kdot1}) will push the positive charged particle away from the y-o-z plane but drive negative charged particles to the y-o-z plane,
enhancing the former $v_2$ but suppressing the latter $v_2$. This is similar to the argument in Ref.~\cite{Sun16}. To test this idea, we break the
correlation between the $z$-component velocity and the $x$ coordinate by exchanging randomly $x$ coordinates in the initial parton list, and
indeed they become uncorrelated as shown by the flat line in the right panel of Fig.~\ref{Fig4}. With this initial parton distribution, it is shown in the left panel of Fig.~\ref{Fig4} that the slope of $v_2(\bar{u})-v_2(u)$ with respect to $A_{ch}$ becomes slightly positive. This shows that the slope is very sensitive to the initial parton distribution. It is also seen that the overall elliptic flow difference as well as its slope with respect to $A_{ch}$ can be affected by
the strength of the real magnetic field as well as the quark-antiquark vector interaction. As expected, the real magnetic field in vacuum
($\sigma_{con}=0$ and $G_{V}=0$), which decays very fast as shown in Fig.~\ref{Fig2}, leads to a negligible slope as shown in the left panel of Fig.~\ref{Fig4}. The space-time evolution of the effective magnetic field from the net quark flux is quite different from that of the real
magnetic field as shown in Fig.~\ref{Fig2}, and it reduces the slope of $v_2(\bar{u})-v_2(u)$ with respect to $A_{ch}$. The solid lines are from the linear fit, and the slopes as well as the fitting errors are listed in Table~\ref{T1}.

\begin{table*}\small
	\centering
	\caption{Slope parameter $r$ from the linear fit of the $v_2$ difference between negative and positive
charged particles as a function of $A_{ch}$ at the freeze-out stage of the partonic phase, the initial stage of the hadronic phase right after hadronization, and the
freeze-out stage of the hadronic phase, for different scenarios of partonic dynamics.}
	\begin{tabular}{|c|c|c|c|c|}
		\hline
		  & $\sigma_{con}>0$, $G_V=0$ & $\sigma_{con}>0$, $G_V=0$, uncorr & $\sigma_{con}=0$, $G_V=0$& $\sigma_{con}>0$, $G_V=1.1G_S$\\
		\hline
		 $r(\%)$ for freeze-out $v_2(\bar{u})-v_2(u)$ & $-3.244 \pm 2.139$ & $1.161\pm 2.073$ & $0.385\pm 2.112$ & $-0.828\pm 1.909$ \\
		\hline
		 $r(\%)$ for initial $v_2(\pi^-)-v_2(\pi^+)$ & $2.488 \pm 2.113$ & $-0.475\pm 2.076$ & $-2.699\pm 2.073$ & $0.819\pm 1.999$ \\
		\hline
		 $r(\%)$ for freeze-out $v_2(\pi^-)-v_2(\pi^+)$  & $0.301 \pm 1.154$ & $0.422\pm 1.139$ & $0.861\pm 1.14$ & $-0.596\pm 1.037$\\
		\hline
	\end{tabular}
	\label{T1}
\end{table*}

\subsection{Charge asymmetry dependence of the elliptic flow splitting in the hadronic phase}

\begin{figure}[ht]
	\includegraphics[scale=0.3]{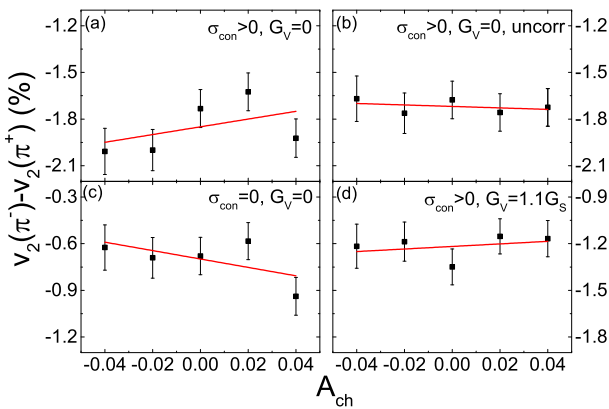}
	\caption{(Color online) Elliptic flow difference between $\pi^{-}$ and $\pi^{+}$
		as a function of the electric charge asymmetry $A_{ch}$ at the initial stage of the hadronic phase right after hadronization in midcentral Au+Au collisions at $\sqrt{s_{NN}}=200$ GeV for different scenarios of partonic dynamics.} \label{Fig5}
\end{figure}

\begin{figure}[ht]
	\includegraphics[scale=0.3]{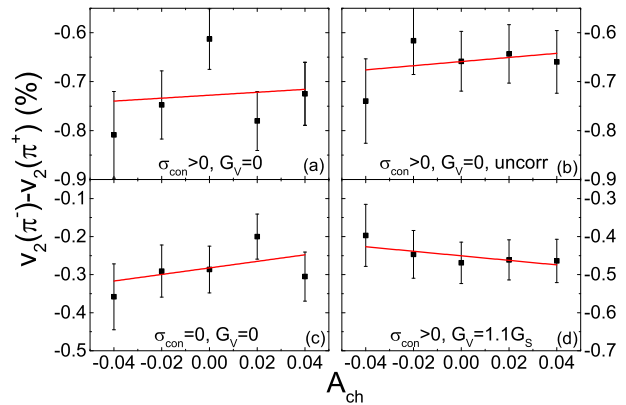}
	\caption{(Color online) Same as Fig.~\ref{Fig5} but at the freeze-out stage of the hadronic phase.} \label{Fig6}
\end{figure}

The $v_2$ splitting of charged partons is not the experimental observable, and can be modified in the hadronization process and after the
hadronic evolution. Also, $A_{ch}$ in the hadronic phase is different from that in the partonic phase, as shown in Fig.~\ref{Fig3}. With a spatial coalescence in the AMPT model, the $v_2$ difference between $\pi^{-}$ and $\pi^{+}$ right after hadronization as well as that after the hadronic evolution as a function of $A_{ch}$ of the corresponding hadronic system are respectively shown in
Fig.~\ref{Fig5} and Fig.~\ref{Fig6}. The slopes of the $v_2$ difference with respect to $A_{ch}$ as well as the fitting errors are listed in Table~\ref{T1}. It is seen
that the overall magnitude of the $v_2$ splitting between $\pi^{-}$ and $\pi^{+}$ is larger than that between $\bar{u}$ and $u$ quarks,
due to the additive effect of $u(\bar{u})$ and $d(\bar{d})$ quarks. The slopes of the $v_2(\pi^-)-v_2(\pi^+)$ with respect to $A_{ch}$ can
already be modified in the hadronization process for different scenarios of partonic dynamics. Due to the violate collisions and decays in the
hadronic evolution, it is seen that not only the overall magnitude of the $v_2$ splitting between $\pi^{-}$ and $\pi^{+}$ but also the
slope of $v_2(\pi^-)-v_2(\pi^+)$ with respect to $A_{ch}$ are changed. Our transport simulations show that the $A_{ch}$ dependence of the
$v_2$ splitting can be largely modified in the hadronization process and during the hadronic evolution. The slopes of final freeze-out $v_2(\pi^-)-v_2(\pi^+)$ with respect to $A_{ch}$ for all scenarios are still much smaller than that observed experimentally~\cite{STAR18}, with the latter as large as about $3\%$. This shows that even if there are some CMW effects in the partonic phase, they are largely modified or washed out in the later stage, and can not lead to the large positive slope of $v_2(\pi^-)-v_2(\pi^+)$ with respect to $A_{ch}$ observed experimentally.

\subsection{Spin polarizations for different quark species}

\begin{figure}[ht]
	\includegraphics[scale=0.3]{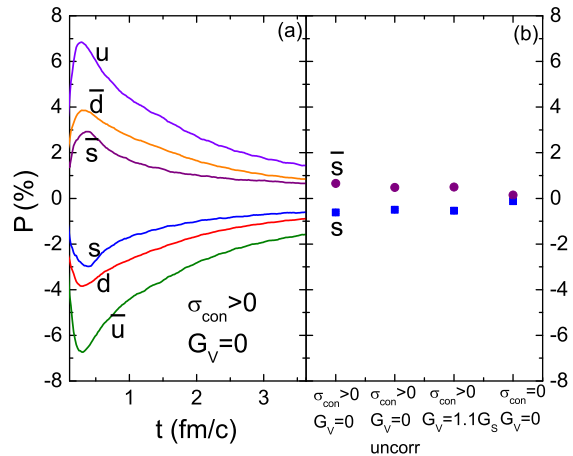}
	\caption{(Color online) Left: Time evolution of the spin polarizations of different quark species for a typical scenario; Right: Spin
polarizations of $s$ and $\bar{s}$ quarks at the freeze-out stage of the partonic phase for different scenarios.} \label{Fig7}
\end{figure}

As seen from Fig.~\ref{Fig2}, the magnetic field is mostly in the $-y$ direction. The spin polarization for partons with respect to the $-y$
direction from the chiral kinetic equations of motion is
\begin{eqnarray} \label{p}
\langle P\rangle=-\frac{\int d^3\vec{r}\frac{d^{3}\vec{k}}{(2\pi)^{3}}c\dot{y}\sqrt{G} f(\vec{r},\vec{k})}{\int
d^3\vec{r}\frac{d^{3}\vec{k}}{(2\pi)^{3}}\sqrt{G} f(\vec{r},\vec{k})}=-\frac{\sum_{n}c_{n}\dot{y}_{n}\sqrt{G_{n}}}{\sum_{n}\sqrt{G_{n}}}.
\end{eqnarray}
In the above, $f(\vec{r},\vec{k})$ is the phase-space distribution function, and in transport simulations the integral is replaced by the
summation, with $c_n$, $\dot{y}_{n}$, and $\sqrt{G_{n}}$ being the corresponding quantities of the $n$th particle. The time evolutions of the
spin polarizations of different quark species under the magnetic field in QGP but without the quark-antiquark vector interaction
($\sigma_{con}>0$ and $G_{V}=0$) are shown in the left panel of Fig.~\ref{Fig7}. It is seen that the spin polarizations first increase and then
decrease with time, as a result of the damping of the magnetic field as shown in Figs.~\ref{Fig1} and \ref{Fig2}. $u(\bar{u})$ quarks have a stronger spin
polarization than $d(\bar{d})$ and $s(\bar{s})$ quarks due to their electric charge difference, while the different spin polarizations between
$d(\bar{d})$ and $s(\bar{s})$ quarks are likely due to their different initial phase-space distributions. $\bar{s}$ quarks have an positive spin polarization compared with $s$ quarks due to their opposite electric charge. If the flow vorticity were further incorporated, the spin polarizations of all quark species are expected to be enhanced, while the splitting of their spin polarizations at freeze-out will remain mostly unchanged, since it is determined by the strength of the magnetic field, as shown in the right panel of Fig.~\ref{Fig7}. It is seen that the uncorrelated initial parton phase-space
distribution or a finite vector coupling constant only slightly modifies the splitting of the spin polarization between $s$ and $\bar{s}$ quarks. According to the STAR
measurement, the difference between the spin polarizations of $\Lambda$ and $\bar{\Lambda}$ is negligible~\cite{STAR18}. Assuming that the
$\Lambda(\bar{\Lambda})$ spin polarization is similar to that of the $s(\bar{s})$ quarks based on the quark coalescence model, the experimentally
observed splitting of the spin polarization between $\Lambda$ and $\bar{\Lambda}$ favors the magnetic field in vacuum, as shown in the right panel of
Fig.~\ref{Fig7}. Such a rapidly decaying magnetic field as shown in Fig.~\ref{Fig2} will further suppress the possible CMW effect as shown in
Figs.~\ref{Fig4}-\ref{Fig6}.

\section{summary and outlook}
\label{summaryhan}

Based on the framework of the extended AMPT model, we have carried out a state-of-the-art transport simulation in order to investigate the
dependence of the elliptic flow splitting between opposite charged particles on the electric charge asymmetry presumedly due to the chiral magnetic wave in midcentral Au+Au collisions at $\sqrt{s_{NN}}=200$ GeV. With the partonic evolution described by the chiral kinetic equations of motion under the magnetic field in QGP mainly generated by spectator protons, a negative slope of the elliptic flow difference between $\bar{u}$ and $u$ quarks with
respect to the electric charge asymmetry is observed. We found that this is due to the correlation between the $z$-component velocity and the $x$
coordinate in the initial parton distribution. The dependence of the elliptic flow difference between $\pi^-$ and $\pi^+$ on the electric charge asymmetry may be affected by the quark-antiquark vector interaction, and is largely modified not only in the hadronization process but also during the hadronic evolution. Neglecting the QGP response to the magnetic field leads to a small/negligible splitting of the spin polarization between $s$ and
$\bar{s}$ quarks, consistent with the similar spin polarization of $\Lambda$ and $\bar{\Lambda}$ observed experimentally, but would further weaken the chiral magnetic wave.

Our study disfavors that the positive slope of the elliptic flow splitting between $\pi^-$ and $\pi^+$ with respect to the electric charge asymmetry,
which is experimentally observed in Ref.~\cite{STAR15}, is due to the chiral magnetic wave. However, the positive slope may be potentially
explained by the mean-field potential effect. It is expected that $u$ and $d$ quarks may be affected by different mean-field potentials in the isospin asymmetric quark matter as a result of the isovector coupling~\cite{Liu16}. Also, $\pi^-$ and $\pi^+$ are affected by different mean-field potentials in the isospin asymmetric hadronic matter~\cite{Xu12}. Their elliptic flow splitting can be generated by their different mean-field potentials since the electric charge asymmetry is also related to the isospin asymmetry of the medium. Such a study is in progress.

\begin{acknowledgments}

We acknowledge helpful discussions with Che Ming Ko,
and thank Chen Zhong for maintaining the high-quality performance of the
computer facility. This work was supported by the Major State Basic
Research Development Program (973 Program) of China under Contract
No. 2015CB856904, and the National Natural Science
Foundation of China under Grant No. 11475243 and No. 11421505.

\end{acknowledgments}

\end{document}